\documentclass[a4paper,11pt,onecolumn,preprint]{elsarticle}
\usepackage[english]{babel}
\usepackage{amsmath}
\usepackage{color}
\usepackage{natbib}
\biboptions{square,numbers,sort&compress}

\journal{Physics Letters B}

\begin{document}
\begin{frontmatter}
\title{Geometric standing wave braneworld and field localization in Lyra manifold} 

\author[ifce]{L. J. S. Sousa}
\ead{luisjose@fisica.ufc.br}

\author[ufca]{J. E. G. Silva}
\ead{euclides.silva@ufca.edu.br}

\author[ifcej]{W. T. Cruz}
\ead{wilami@ifce.edu.br}

\author[ufc]{C. A. S. Almeida}
\ead{carlos@fisica.ufc.br}

\address[ifce]{Instituto Federal de Educa\c{c}\~{a}o Ci\^{e}ncia e Tecnologia do Cear\'{a}\\ (IFCE) - Campus de Maracana\'{u} Av. Contorno Norte, 10 - Distrito Industrial, Maracana\'{u}, 61936-000 - Cear\'{a} - Brazil}

\address[ufca]{Universidade Federal do Cariri (UFCA) - Centro de Ciências e Tecnologia. \\
Av. Tenente Raimundo Rocha S/N, Cidade universit\'{a}ria \\ 63048-080 Juazeiro do Norte - Cear\'{a} - Brazil}

\address[ifcej]{Instituto Federal de Educa\c{c}\~ao, Ci\^encia e Tecnologia do Cear\'a (IFCE), Campus
Juazeiro do Norte - 63040-540 Juazeiro do Norte, CE, Brazil}

\address[ufc]{Departamento de F\'{\i}sica - Universidade Federal do Cear\'{a} - UFC \\  C.P. 6030, 60455-760 Fortaleza - Cear\'{a} - Brazil
}


\begin{abstract}

In this work, we propose a standing wave braneworld based on Lyra geometry scenario. The Lyra displacement vector provides a modification in Einstein equations which can be interpreted as a noninteracting phantom scalar. From the Einstein's equation in Lyra manifold,  a 5D standing wave braneworld is constructed in the presence of a cosmological constant. Unlike other standing wave solutions presented in the literature, no matter field is necessary to obtain these new solutions. We analyse the properties of the scalar, gauge vector and fermionic fields in this model, highlighting the relevance of geometric structure in the process of trapping matter.  
 
\end{abstract}


\end{frontmatter}

\section{Introduction}

Since its appearance in theoretical physics literature, several braneworld models have been proposed. Some of them must be cited by its particular importance in the context of braneworld theories, mainly by its capacity to present a solution to fundamental Standard Model problems. Answers to the hierarchy problem were proposed on the works \cite{Randal1999} and \cite{Arkani2001}; the origin of dark matter was proposed in \cite{Arkani2000}; solutions to the cosmological constant problem and cosmic acceleration were presented in the works \cite{Chen2000} and \cite{Khoury2001}, respectively.    

All braneworld models carry the idea that our world is a brane (or membrane) embedded in a higher dimensional world and all Standard Model field are trapped on the brane, while gravity is free to propagate in the bulk. These assumptions are important since the model has to  match experimental particle and gravitational results \cite{Tao2017}. On the other hand, this assumption brings the problem of how are the trapping mechanisms that assure the localization of the various matter field on the brane. Several  mechanisms are presented in literature assuring  localization of scalar, fermionic and vector field \cite{Oda2000a, Oda2000b, Gherghetta2000, Giovannini2001, Kehagias2001, nosso4, nosso5, Torrealba2010, Gogberashvili2011, Silva2011, Gogberashvili2012, Sousa2012, Merab2012}. Particularly, for a class of geometrical braneworld, called Weyl brane, the zero mode of vector field cannot be localized implying the necessity of new localization mechanisms, like the interaction of vector field and Weyl scalar \cite{Tao2017}.

In five-dimensional thick brane models,  there is a widely adopted gauge field localization mechanism that consists of coupling the dilaton field directly with the Maxwell kinetic term on the Lagrangian density \cite{Kehagias2001, nosso4}.  The same approach was also used to ensure the localization of the Kalb-Ramond field as shown in the works of Ref. \cite{nosso5}. Based on these results, we are motivated to search for braneworld scenarios where the localization of vector fields can be obtained naturally without extra couplings.

In this work, we construct a standing wave braneworld model from the geometrical structure of Lyra manifold. It consists in an anisotropic brane like the original model one \cite{Merab2009}, but with the fundamental difference that in our case, there is no needing of a scalar (phantom) matter field to interact with gravity and thus generate the model. Indeed, the geometrical aspects of Lyra geometry gives the standing wave braneworld solution. On the other hand, in this work it is still considered the localization of various zero mode matter fields showing, mainly, that the vector field is localized without the necessity of extra interactions. This work is divided as follows: in section II a review of Lyra geometry is given where it is shown that the Einstein's equations in Lyra geometry, in normal gauge, is similar to the Einstein's equation in a Riemannian manifold in the presence of a noninteracting scalar phantom field. In section III a standing wave braneworld is obtained from the Einstein's equations in Lyra geometry and, finally, in section IV, the localization of scalar, fermionic and vector zero mode fields are considered.

\section{Lyra geometry: briefly review}

Einstein introduced the notion of geometrization in physics by the identification of gravity potential with the metric tensor of a Riemannian manifold. A connection which characterizes this kind of manifold is metric preserving and torsion free. Hermann Weyl, intending to construct a unified theory of electromagnetism and gravity in the same space-time geometry, proposed a modification of Einstein's general relativity theory. The model introduced by Weyl is a modification of Riemannian manifold, once its connection is torsion free, but it is not metric preserving. Einstein observed that  in this case the atoms spectral frequencies would depend on its past histories. In other words, in such a theory the spectra frequencies of atoms would not have any absolute significance \cite{Sen-1972}. It was in this context that the so-called  `Lyra geometry' theory was first proposed by G. Lyra, as narrated by Sen \cite{Sen-1971}. The Weyl theory was modified by Lyra who constructed a method which is metric preserving but not torsion-free. Weyl and Lyra original works were later generalized in a single method \cite{Sen-1971, Sen-1972}. It is this version of Lira theory that will be revisited in this section.

Lyra introduced a gauge function in the ``structureless manifold'' which results in a modification of the displacement  vector  between two  points, $P(x^M)$ and $P'(x^M + dx^M)$, which, in this case, has the components $V^M = fdx^M$, where $f(x)$ is a nonzero gauge function \cite{Sen-1971}. The parallel transfer of a vector in Lyra's geometry is
\begin{equation}
\delta V^M = -\widetilde{\Gamma}_{AB}^{M}V^Afdx^B ,
\end{equation}
where
\begin{equation}
\widetilde{\Gamma}^{M}_{AB} = \Gamma_{AB}^M - \frac{1}{2}\delta_A^M\phi_B.
\end{equation}
Though $\widetilde{\Gamma}_{AB}$ is asymmetric, the $\Gamma_{AB}^M = \Gamma_{BA}^M$ is symmetric. The $\phi_B$ is a displacement vector which appears in theory as a consequence of the introduction of the gauge function in the structureless manifold \cite{HALFORD-1970}.

The metric or measure of the length of the displacement vector between two points and the integrability of parallel transfer of length are  given in Lyra geometry, respectively, by
\begin{equation}\label{metric-lyra}
ds^2 = g_{MN}fdx^Mfdx^N,
\end{equation}
and
\begin{equation}\label{integrability}
\delta (g_{MN}V^MV^N) = 0.
\end{equation}
From Eq. (\ref{metric-lyra}) and Eq. (\ref{integrability}) a torsion-free Lyra connection is given by \cite{Sen-1971},

\begin{equation}
\Gamma_{AM}^M = f^{-1}\lbrace_{AM}^M\rbrace + \frac{1}{2}(\delta_A^M\phi_B + \delta_B^M\phi_A - g_{AB}\phi^M),
\end{equation}
 where $\lbrace_{AM}^M\rbrace$ are  Christoffel symbols of
the second kind. The curvature scalar derived from this connection is \cite{Sen-1971},
\begin{equation}\label{k-geral}
K = \frac{R}{f^2} + \frac{1}{4}(N - 2)(N - 1)\phi^{A}\phi_{A} + \frac{1}{4}(N - 2)(N - 1)\phi^{A}\hat{\phi}_{A}  + \frac{(N - 1)}{f}\phi_{;A}^{A},
\end{equation} 
where $N$ is the dimension of the bulk, and the capital letters represent the indexes in bulk. Additionally, $\hat{\phi}_\alpha = f^{-1}(\ln f^2 ),\alpha$. The ordinary partial derivative is represented with a comma, $\partial_{\alpha}f = f,\alpha$  the covariant derivative  with a semicolon, $\phi_{;\alpha}^{\alpha} = \nabla_{\alpha}\phi^{\alpha}$.


The Einstein equations in Lyra manifold are derived applying the variational approach to the scalar curvature Eq. (\ref{k-geral}) by two kinds of procedures. In the first case, we assume the standard gauge ansatz, $f = 1$, which implies $\hat{\phi}_A = 0$. 

The scalar curvature and Einstein's equations, in this case, are respectively 
\begin{equation}\label{k-gauge1}
K = R + \frac{1}{4}(N - 2)(N - 1)\phi^{A}\phi_{A}   + (N - 1)\phi_{;A}^{A}
\end{equation} 
and
 \begin{equation} \label{einstein1}
  G_{A B} + \frac{1}{4}(N - 2)(N - 1)(\phi_{A} \phi_{B} -\frac{1}{2}g_{AB} \phi_{C} \phi^{C}) = 0.
  \end{equation}
If we write $\phi_A$ as a derivative of a scalar function, $\phi_A = \partial_A h$, where $h = h(x)$ is a scalar function, the Einstein's equation Eq. (\ref{einstein1}) results

\begin{equation} \label{einstein11}
  R_{A B} - \frac{1}{2}g_{AB}R =  \frac{1}{4}(N - 2)(N - 1)T_{AB}^{phanton},
  \end{equation}
where $T_{AB}^{phantom} = -\partial_A h\partial_B h + \frac{1}{2}g_{AB} \partial_C h \partial^C h$ is the energy-momentum tensor of a scalar ghost field, without interaction. In reference \cite{HALFORD-1970},  the particular case  $N = 4$ for Eq. (\ref{k-gauge1}) and  Eq. (\ref{einstein1}) may be found. 

Another procedure to achieve the Einstein equations in Lyra manifold consists in applying the variational method to the general scalar curvature Eq. (\ref{k-geral}):  
\begin{equation}\label{prin-var}
\delta\int K(\sqrt{-g})(f dx^1 f dx^2\cdot \cdot \cdot f dx^N) = \delta\int f^N K(\sqrt{-g})dx^1\cdot \cdot \cdot dx^N = 0.
\end{equation}
%
%
%
After some calculations the result is
\begin{eqnarray}\label{Eins7}
G^{AB} + \frac{f^2}{4}(N - 2)(N - 1) (\phi^{A} \phi^{B} + \hat{\phi}^{A} \phi^{\beta})\\\nonumber - \frac{f^2}{8}(N - 2)(N - 1) g^{AB}(\phi_{C} \phi^{C}  + \hat{\phi}_{C} \phi^{C}) = 0,
\end{eqnarray} 
\begin{equation}\label{Eins8}
 3\phi^{A} + \frac{3}{2}\hat{\phi}^{A} = 0.
 \end{equation}
If we insert Eq. (\ref{Eins8}) in  Eq. (\ref{Eins7}) we find
\begin{equation}\label{Eins2}
G_{A B} - \frac{f^2}{16}(N - 2)(N - 1) \hat{\phi}_{A} \hat{\phi}_{B} + \frac{f^2}{32}(N - 2)(N - 1) g^{\alpha \beta} \hat{\phi}_{C} \hat{\phi}^{C})  = 0.
\end{equation} 
In terms of the gauge function $f$, the Einstein's equation results:
\begin{equation}\label{Eins14}
G_{A B} - \frac{f^{-2}}{4}(N - 2)(N - 1) \partial_{A }f \partial_{B}f + \frac{f^{-2}}{8}(N - 2)(N - 1) g_{A B } \partial_{M}f \partial^{M}f  = 0.
\end{equation} 
Once again it is possible to write the right hand side (RHS) \ of Eq. (\ref{Eins14}) as a multiple of a scalar field energy-momentum (a real scalar field in this case)
\begin{equation}\label{Eins225}
R_{A B} - \frac{1}{2}g_{AB}R =  \frac{f^{-2}}{4}(N - 2)(N - 1)T_{AB},
\end{equation}
where $T_{AB} = \partial_{A }f \partial_{B}f - \frac{1}{2} g_{A B } \partial_{M}f \partial^{M}f  $ is  the energy-momentum tensor of the scalar field $f$ for a free particle. Finally, inserting $\hat{\phi}_A = -2\phi_A$, from Eq. (\ref{Eins8}), in Eq. (\ref{Eins7}), the resultant equation is 
\begin{equation}\label{Eins2234}
R_{A B} - \frac{1}{2}g_{AB}R =  \frac{f^2}{4}(N - 2)(N - 1)(\phi_A \phi_B - \frac{1}{2} g_{A B } \phi_C \phi^C) ,
\end{equation}
which may be written as a gravity-scalar field equation, by assuming $\phi_A$  as a derivative of a scalar function. But in this case, unlike usual gauge result, it would be a real scalar field. 

Equations Eq. (\ref{Eins225}) were derived in \cite{Sen-1971} for the case $N = 4$. Several solutions in four dimensional space-time were obtained where different possible interpretations for the Lyra displacement vector were given  \cite{Casana2005, Gad2015, Darabi2015, Shchigolev2015, Saadat2016, Shchigolev2017}. Solutions in higer dimensional theory were considered in recent years in different contexts  \cite{Rahaman2001,Mohanty2009}. In the next section  we will look for solutions of Einstein's equation (\ref{einstein1}), in the presence of a cosmological constant. 

\section{ Standing wave solution of Einstein-Lyra equations}
In the presence of a cosmological constant, Eq. (\ref{einstein11}) will result 
\begin{equation} \label{einstein17}
  R_{A B} - \frac{1}{2}g_{AB}R =  \frac{1}{4}(N - 2)(N - 1)T_{AB}^{phanton} - \Lambda g_{AB},
  \end{equation}
where $T_{AB}^{phantom} = -\partial_A h\partial_B h + \frac{1}{2}g_{AB} \partial_C h \partial^C h$ is the energy-momentum tensor of a scalar ghost field, without interaction, and $\Lambda$ is a cosmological constant. We consider a general anisotropic metric ansatz with a warp factor to find a 5D Standing wave braneworld solutions to the system of equations given above. Thus, we have
\begin{equation} \label{metric-aniso5d}
ds^{2}= e^{A(r)}\left( dt^{2} - e^{u(r, t)}dx^{2} - e^{u(r, t)}dy^{2} - e^{-2u(r, t)}dz^{2} \right) - dr^{2}, 
\end{equation}
where $A(r)$ depends only on the extra dimension $r$ and $u = u(r,t)$ is a function of  $t$ and $r$. For $N = 5$ Eq. (\ref{einstein17}) may be written as  
\begin{equation}\label{eq.19}
R_{AB} - \frac{1}{2}g_{AB}R = 3\left(-\partial_Ah\partial_Bh + \frac{1}{2} g_{AB}\partial^Ch\partial_Ch\right) - \Lambda g_{AB}.
\end{equation} 
In the next, the factor `3', will be redefined to insert a 5D gravitational constant, $\kappa_5$, resulting for the Einstein's equation (\ref{eq.19}),

\begin{equation}
R_{AB} - \frac{1}{2}g_{AB}R = \kappa_5^2\left(-\partial_Ah\partial_Bh + \frac{1}{2} g_{AB}\partial^Ch\partial_Ch\right) - g_{AB}\Lambda.
\end{equation} 
If we eliminate $R$ from equation above, the Ricci tensor may be written as
\begin{equation} \label{eq.76}
R_{AB} = -\kappa_5^2 \partial_Ah\partial_Bh + \frac{2}{3}g_{AB}\Lambda.
\end{equation} 
Assuming dot (.) as time derivaive and line (') as r derivative, for the metric Eq. (\ref{metric-aniso5d}), the non zero components of the Ricci tensor are 
\begin{equation} \label{Ricci-aniso-xx5d}
R_{xx} = R_{yy} = -\frac{1}{2} e^{A + u} \left( 2 A^{'2} +  A^{''} -e^{-A}\ddot{u} +2 A^{'} u^{'} + u^{''}\right),
\end{equation}
\begin{equation} \label{Ricci-aniso-zz5d}
R_{zz} = -\frac{1}{2} e^{A -2 u} \left( 2 A^{'2} +  A^{''} + 2 e^{-A}\ddot{u} -4 A^{'} u^{'} -2 u^{''} \right),
\end{equation}
\begin{equation} \label{Ricci-aniso-tt5d}
R_{tt} = \frac{1}{2} e^{A}\left( 2A^{'2} +A^{''} -3 e^{-A}\dot{u}^2 \right),
\end{equation}
\begin{equation} \label{Ricci-aniso-rt5d}
R_{rt} = -\frac{3}{2}\dot{u}u^{'},
\end{equation}
\begin{equation} \label{Ricci-aniso-rr5d}
R_{rr} = \frac{1}{2}\left(- 2A^{'2} - 4 A^{''}  - 3u^{'2}\right).
\end{equation}
The RHS of Eq. (\ref{eq.76})  are
\begin{equation} \label{Ricci-xx2}
R_{xx} = -\frac{2}{3} e^{A + u}\Lambda = R_{yy},
\end{equation}
\begin{equation} \label{Ricci-zz2}
R_{zz} = -\frac{2}{3} e^{A -2u}\Lambda,
\end{equation}
\begin{equation} \label{Ricci-tt2}
R_{tt} =  -\kappa _{5} ^{2} \dot{h}^2 + \frac{2}{3} \Lambda e^{A},
\end{equation}
\begin{equation} \label{Ricci-rt2}
R_{rt} = -\kappa _{5} ^{2} \dot{h}h^{'},
\end{equation}
\begin{equation} \label{Ricci-rr2}
R_{rr} = -\kappa _{5} ^{2} h^{'2} - \frac{2}{3}\Lambda.
\end{equation}
So, the system of differential equations that has to be solved is
\begin{equation} \label{eq.82}
 -\frac{1}{2} e^{A + u} \left( 2 A^{'2} +  A^{''} -e^{-A}\ddot{u} +2 A^{'} u^{'} + u^{''}\right) = - \frac{2}{3} e^{A + u}\Lambda,
\end{equation}

\begin{equation} \label{eq.83}
 -\frac{1}{2} e^{A -2 u} \left( 2 A^{'2} +  A^{''} + 2 e^{-A}\ddot{u} -4 A^{'} u^{'} -2 u^{''} \right) = - \frac{2}{3} e^{A -2u}\Lambda,
\end{equation}
\begin{equation} \label{eq.84}
\frac{1}{2} e^{A}\left( 2A^{'2} +A^{''} -3 e^{-A}\dot{u}^2 \right) =  -\kappa _{5} ^{2} \dot{h}^2 + \frac{2}{3} e^{A} \Lambda ,
\end{equation}
\begin{equation} \label{eq.85}
-\frac{3}{2}\dot{u}u^{'} = -\kappa _{5} ^{2} \dot{h}h^{'},
\end{equation}
\begin{equation} \label{eq.86}
\frac{1}{2}\left(- 2A^{'2} - 4 A^{''}  - 3u^{'2}\right) = -\kappa _{5} ^{2} h^{'2} - \frac{2}{3}\Lambda.
\end{equation}
As can be seen directly, Eq. (\ref{eq.85}) implies $\dot{h} = \sqrt{\frac{3}{2\kappa _{5} ^{2}}}\dot{u}$ and $h' = \sqrt{\frac{3}{2\kappa _{5} ^{2}}}u'$. If we choose $A''(r) = 0$ and $A'^2(r) = \frac{2}{3}  \Lambda$, which implies $\Lambda > 0$, the equation for $u(r, t)$ results
\begin{equation}
-e^{-A}\ddot{u} +2 A^{'} u^{'} + u^{''} = 0.
\end{equation}
This equation may be separated by choosing  $u(r,t) = Q(r)\sin(\omega t)$ resulting
\begin{equation}
Q''(r) + 2A'(r)Q'(r) + \omega^2e^{-A(r)}Q(r) = 0.
\end{equation}
For  $A(r) =  \sqrt{\frac{2}{3}\Lambda}|r|$ the general solution to this equation is
\begin{equation}
Q(r) = C_1e^{-\sqrt{\frac{2}{3}\Lambda}|r|}J_2\left(\sqrt{\frac{6}{\Lambda}}\omega e^{-\sqrt{\frac{\Lambda}{6}}|r|}\right) + C_2e^{-\sqrt{\frac{2}{3}\Lambda}|r|}Y_2\left(\sqrt{\frac{6}{\Lambda}}\omega e^{-\sqrt{\frac{\Lambda}{6}}|r|}\right),
\end{equation}
\noindent where $J_2, Y_2$ are first and second kind Bessel function of order 2, respectively, and $C_1$, $C_2$ are integration constants. This is the same solution found in Refs. \cite{Merab2009, Merab2011}. Similar solutions in six dimensions were found in \cite{Sousa2012, Sousa2014}. For the next we set $C_2 = 0$ to compare our field localization results with the work \cite{Merab2012}. 

Althoug the exponential dependence prevent the second order Bessel funtion of second kind, $Y_2$, to diverge in origin, we will consider only Bessel function of the first kind $Y_1$, by setting $C_2 = 0$
\begin{equation}\label{disp-vec-partic-sol}
Q(r) = C_1e^{-\sqrt{\frac{2}{3}\Lambda}|r|}J_2\left(\sqrt{\frac{6}{\Lambda}}\omega e^{-\sqrt{\frac{\Lambda}{6}}|r|}\right) .
\end{equation}
The condition $Q(0) = 0$ will result in a corresponding quantization condition for the oscilation frequencies of oscilations 
\begin{equation}\label{Bessel-zeros}
\omega \sqrt{\frac{6}{\Lambda}} = X_{2,n},
\end{equation}
where $X_{2,n}$ is the n-st root of the Bessel function $J_2$, \cite{Merab2009}. Solution of the Eq. (\ref{disp-vec-partic-sol}) can be interpreted as standing wave propagating to the right. For  an increasing warp factor, $A(r) > 0$, the number of zeros, Eq. (\ref{Bessel-zeros}), is finite and can be restricted to one zero in the origin by assuming $\omega \sqrt{\frac{6}{\Lambda}} = X_{1} = 5.14$ \cite{Merab2012}.

\begin{figure}
\label{radialfielddependence}
\begin{center}
\includegraphics[scale=0.5]{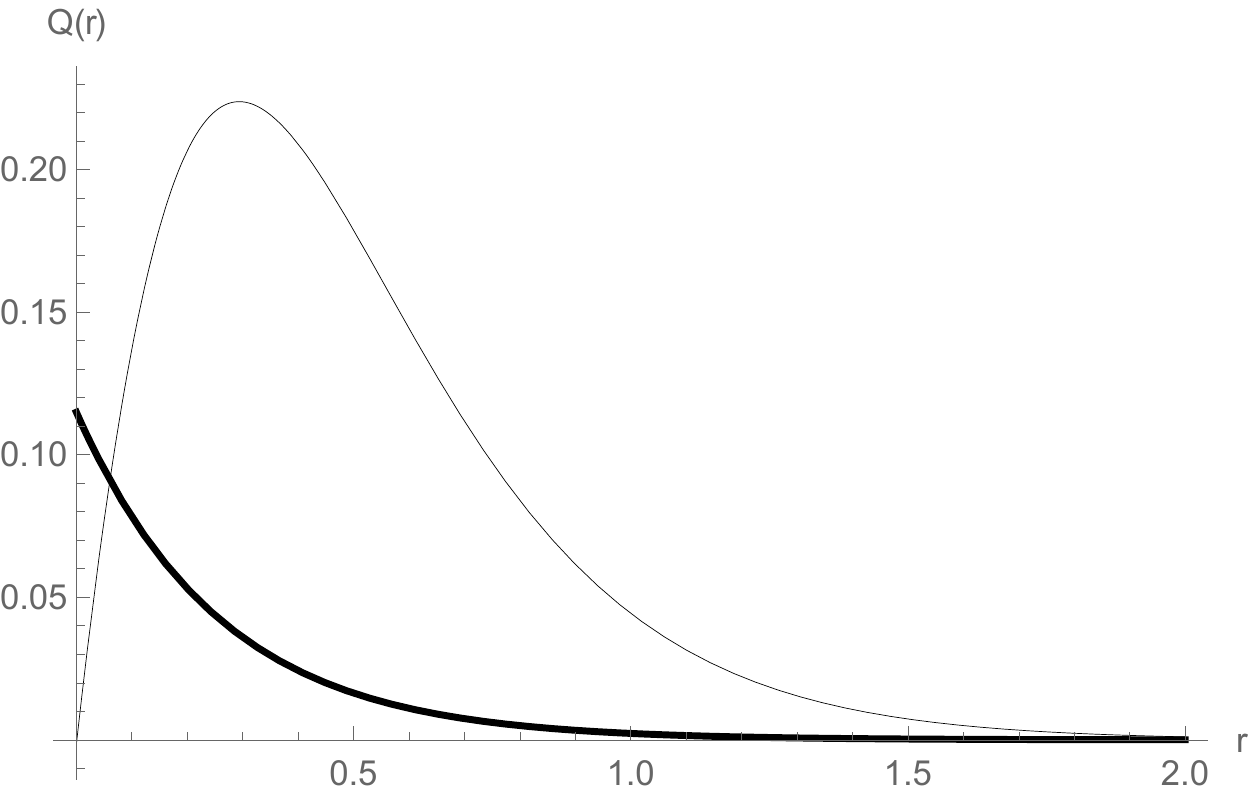}
\caption{Profile of the Lyra scalar for a flat Lorentzian brane (thin line) and for an anisotropic brane (thick line).}
\end{center}
\end{figure}
We plotted the function $Q(r)$ in Fig. (1) for $\sqrt{\frac{6}{\Lambda}}\omega = 5.14$ (thin line and for $\sqrt{\frac{6}{\Lambda}}\omega = 1$ (thick line). Since the Lyra scalar field $h(r,t)=\sqrt{\frac{3}{2\kappa_5}}Q(r)\sin(\omega t)$, the configurations where $Q(0)=0$ represents a flat Lorentz invariant standing wave branes with a vanishing phantom field.
In the next secton we will consider localizaton of matter fields on the brane. The equations for this will contain oscilatory exponents a little bit dificult to treat algebrically. Some simplification can be obtained if one consider that the frequency of particles of matter field is very small in comparison with the standing wave frequency $\omega$. The oscilatory terms in matter field equations may be averaging and some important results to be used in the next section can be found below \cite{Merab2012} 

\begin{equation}\label{average-u}
\langle u \rangle = \langle \dot{u} \rangle = \langle u' \rangle = \langle \dot{\left( e^{bu} \right)} \rangle = 0,
\end{equation}
where $b$ is a constant, dot and prime mean, respectively, derivatives with respect time and extra dimension $r$. The non vanishing expressions are given below 
\begin{equation}\label{average-exp}
\langle  e^{bu}  \rangle = I_0,
\end{equation}
\begin{equation}\label{average-deriv-exp}
\langle \left( e^{bu} \right)' \rangle = b^2ZZ'\left[I_0 + \frac{\pi}{2}  \left(I_0L_1 - I_1L_0\right)\right].
\end{equation}
The terms between brackets are modified Bessel, $I_0,\, I_1$, and Struve $L_0,\, L_1$ functions for the arguments $bZ$.

\section{Field localization} \label{section3}
\label{vetor-local}

\subsection{Scalar field}

In order to study the behaviour of scalar field in Lyra standing-wave braneworld we consider the action for a minimally coupled scalar
\begin{equation} \label{scalar-action}
S_{\Phi}=\frac{1}{2}\int d^Nx \sqrt{-g} g^{MN} \partial_M \Phi \partial_N\Phi.
\end{equation}
The equations of motion to this action are obtained from the condition
\begin{equation}
\partial_M \left( \sqrt{-g} g^{MN}  \partial_N\Phi \right) = 0,
\end{equation}
To investigate the localization of the scalar field zero mode we set $\sqrt{\frac{2}{3}\Lambda}r = 2ar$ and $r > 0$, resulting the following equations
\begin{equation} \label{move.eq}
\left[\partial_t^2 - e^{-u}(\partial_x^2 + \partial_y^2) - e^{2u}\partial_z^2\right]\Phi = e^{2ar}.\left(e^{4ar} \Phi' \right)'
\end{equation}
To separate the brane dimensions from the extra dimension one assumes  
\begin{equation}
\Phi(x^{\nu},r) = e^{ip_{\nu}x^{\nu}}\xi(r).
\end{equation}
With this chosen, the equation (\ref{move.eq}) results
\begin{equation}
\left(e^{4ar} \xi' \right)' = e^{2ar}\left[(p_x^2 + p_y^2)e^{-u} + p_z^2e^{2u} - p_t^2 \right]\xi.
\end{equation} 
It is still difficult to solve this equation, but in the case that the standing wave frequencies are much larger than energy frequencies of particles in the brane, it is possible to replace the oscillatory exponent by their time average which gives us a $r$ dependent variable. 
We will then have

\begin{equation} \label{move.eq2}
\left(e^{4ar} \xi' \right)' - e^{2ar}P^2(r)\xi = 0,
\end{equation}
where 
\begin{equation}\label{eq-for-P}
P^2(r) = (\langle e^{-u}\rangle - 1)(p_x^2 + p_y^2) + (\langle e^{2u}\rangle - 1)p_z^2.
\end{equation}
The parameters $p_{\nu}$ may be interpreted as energy-momentum components along the brane, which obey the dispersion relation
\begin{equation}
p_t^2 - (p_x^2 + p_y^2 + p_z^2) = 0.
\end{equation}
It is not possible to find an analytic solution to Eq. (\ref{move.eq2}). But it is possible to investigate what happens to the wave function in the regions $r \rightarrow 0 $ and $r \rightarrow \infty$. For this case
\begin{equation}
P^2(r)\vert_{r\rightarrow 0} \sim r^2,
\end{equation} 
\begin{equation}
P^2(r)\vert_{r\rightarrow \infty} const
\end{equation} 
and the $r-dependent$ variable
\begin{equation}
\xi(r)\vert_{r\rightarrow 0} \sim constant,
\end{equation} 
\begin{equation}
\xi(r)\vert_{r\rightarrow \infty} \sim e^{-4ar}.
\end{equation} 
We conclude that in the limit $r \rightarrow \infty$ the wave function assumes a decreasing exponential form what assures the localization of the scalar zero mode on the brane. 

\subsubsection{Nonminimal coupling}

If one consider the action for the scalar field with a nonminimal coupling the resultant scalar field equation reads

\begin{equation}\label{move.eq.coupling}
\partial_M \left( \sqrt{g} g^{MN}  \partial_N\Phi \right) = \epsilon \kappa_5^2\left( \frac{5}{3}\Lambda - e^{-2ar}\dot{h}^2 + h'^2\right)\Phi,
\end{equation}

For the extra dimension one find

\begin{equation} \label{move.r-couplimg}
\left(e^{4ar} \xi' \right)' - e^{2ar}\bar{P}^2(r)\xi = 0,
\end{equation}
where 
\begin{equation}\label{eq-for-P-bar}
\bar{P}^2(r) = (\langle e^{-u}\rangle - 1)(p_x^2 + p_y^2) + (\langle e^{2u}\rangle - 1)p_z^2 - \frac{5}{3}\Lambda\epsilon\kappa_5^2e^{-2ar}.
\end{equation}

We see that Eq. (\ref{eq-for-P-bar}) differs from Eq. (\ref{eq-for-P}) by the term $\frac{5}{3}\Lambda\epsilon\kappa_5^2e^{-2ar}$. For $r \rightarrow 0$ this term contributes with a constant in comparison with the same approximation obtained in Eq. (\ref{eq-for-P}). For $r \rightarrow \infty$ there is no diference from the results obtained for      
Eq. (\ref{move.eq2}). On the other hand, Eq. (\ref{move.eq.coupling}) shows that the extra term in the equation of motion, coming from the nominimal coupling, may be interpreted as a time dependent mass. So, this time dependent mass add a constant in the r-dependent variable solution. This result sugests that  a time dependent mass may play a relevant role in the mecanism of field localization.

\subsection{Gauge field}

The procedure to analyze the zero mode localization for the gauge field is similar to the one used in the last subsection. However,  the calculations are just a bit more difficult for this case.

We begin by considering the 5D action for the gauge vector field
\begin{equation}
S = - \frac{1}{4}\int d^5x\sqrt{g}g^{MN}g^{PR}F_{MP}F_{NR},
\end{equation}
where $F_{MP} = \nabla_M^{L} A_P - \nabla_P^{L} A_M$. By using the torsion-free connection, the coupling term between the gauge field $A_M$ and the Lyra vector $\Phi_A$ vanishes and the field strength has the usual form $F_{MP} = \partial_M A_P - \partial_P A_M$.
Now we consider the system of equations 
\begin{equation} \label{gauge-move-eq}
\frac{1}{\sqrt{g}}\partial_M\left(\sqrt{g}g^{MN}g^{PR}F_{NR} \right) = 0.
\end{equation}
By assuming that the components of the vector field $A_{M}$ are given as follows
\begin{eqnarray}\label{vector-comp}
& & A_t(x^M) = v(r)a_t(x^{\mu}),\nonumber\\
& & A_x(x^M) = e^{u(t,r)}v(r)a_x(x^{\mu}), \nonumber\\
& & A_y(x^M) = e^{u(t,r)}v(r)a_y(x^{\mu}), \nonumber\\
& & A_z(x^M) = e^{-2u(t,r)}v(r)a_z(x^{\mu}),\nonumber\\
& & A_r(x^M) = 0,
\end{eqnarray}
where the functions $a_{\mu}(x^{\mu})$ are the brane components of $A_M$ and the scalar factor $a(r)$ depends only on the extra dimension $r$, one simplify the system of Eqs. (\ref{gauge-move-eq}). First we observe that the time averaging of the $r$ component of Eq. (\ref{gauge-move-eq}) results

\begin{equation}
\partial_{\alpha}\left(g^{\alpha \beta}A'_{\beta} \right) = 0.
\end{equation} 
This gives the condition
\begin{equation}\label{gauge-cond}
g^{\alpha\beta}\partial_{\alpha}A_{\beta} = \eta^{\alpha\beta}\partial_{\alpha}a_{\beta} = 0. 
\end{equation} 

The vector components chosen above, Eq. (\ref{vector-comp}) and Eq. (\ref{gauge-cond}), represent the complete  set of gauge conditions.
The other four equations in the system (\ref{gauge-move-eq}) are given by
\begin{equation}
\partial_{\mu}\left(g^{\mu\nu}g^{\alpha\beta}F_{\nu\beta} \right) - \frac{1}{\sqrt{g}}\left(\sqrt{g}g^{\alpha\beta}A'_{\alpha} \right)' = 0.
\end{equation}   
The time averaging version of this equation together with the gauge conditions in Eq. (\ref{gauge-cond}), give 
\begin{equation}\label{gauge-aver}
vg^{\alpha\beta}\partial_{\alpha}\partial_{\beta}a_{\mu} + e^{-2a|r|}\left(e^{-2a|r|}v' \right)'a_{\mu} = 0.
\end{equation} 
In order to assure zero mode localization on the brane one requires that the components of 4D vector on the brane assume
\begin{equation}
a_{\mu}(x^{\nu})  \sim \epsilon_{\mu}e^{ip_{\nu}x^{\nu}},
\end{equation}  
where $p_{\nu}$ represent the components of energy-momentum tensor on the brane. With this considerations, the Eq. (\ref{gauge-aver}) for the scalar factor $v(r)$ results
\begin{equation}
\left(e^{2a|r|}v' \right)' - P^2(r)v = 0,
\end{equation}
where $P(r)$ is given in  Eq. (\ref{eq-for-P}). Once again it is not necessary to obtain the general solution for the variable $v(r)$ but only its asymptotic value near and far from the brane. These values are, respectively:
  
\begin{equation}
v(r)|_{r\rightarrow0} \sim constant,
\end{equation}

\begin{equation}
v(r)|_{r\rightarrow \infty} \sim e^{-2a|r|}	.
\end{equation}
 This result show  that the wave function for the gauge field zero mode has a maximum on the brane and falls off far from the brane. On this way, we can assure the zero mode localization.

\subsection{Localization of the fermionic field} \label{section4}
%
%
%
 The fermionic field action can be written in a five dimension scenario as
%
\begin{equation} \label{fermion-action}
S = \int {d^{5} x \sqrt{-g} \bar{\Psi} i \Gamma ^{M} D_{M} \Psi},
\end{equation}
where  $\Gamma ^{M}$'s represent the Dirac matrices in a curved spacetime which, in  turn,  are related to Dirac matrices in flat spacetime as
\begin{equation} \label{gam-matr}
\Gamma ^{M} = e_{\bar{M}}^{M} \gamma ^{\bar{M}},
\end{equation}
\noindent where the  \textit{vielbein} $h_{\bar{M}}^{M}$ is defined by the relation
\begin{equation} \label{vielbein}
g_{M N} = \eta _{\bar{M} \bar{N}}e_{M}^{\bar{M}} e_{N}^{\bar{N}}.
\end{equation}

\noindent Finally, the covariant derivative has the standard form
\begin{equation} \label{deri-covar}
D_{M} = \partial _{M} + \frac{1}{4} \Omega_{M} ^{\bar{M} \bar{N}} \gamma _{\bar{M}} \gamma _{\bar{N}},
\end{equation}

\noindent where   $\Omega_{M} ^{\bar{M} \bar{N}}$ is the spin connection.

For the metric on Eq. (\ref{metric-aniso5d}), the relation between Dirac matrices in curved and flat spacetime, Eq. (\ref{gam-matr}), are
\begin{equation*}
\Gamma^t = e^{-a|r|}\gamma^t,   
\end{equation*}  
\begin{equation*}
\Gamma^x = e^{-a|r| - u/2}\gamma^x,   
\end{equation*}  
\begin{equation}\label{gamma-matr}
\Gamma^y = e^{-a|r| - u/2}\gamma^y,   
\end{equation}  
\begin{equation*}
\Gamma^z = e^{-a|r| + u}\gamma^t,   
\end{equation*}  
\begin{equation*}
\Gamma^r = i\gamma^r.
\end{equation*}  
\noindent In the same background, the spin connection has the following nonvanishing components 
\begin{eqnarray}
\label{spin-conx-nonvan}
\lefteqn{ \Omega_{t} ^{\bar{t} \bar{r}} =  -\left( e^{a|r|}\right)';  \hspace{5 pt} \Omega_{x } ^{\bar{x} \bar{r }} = \Omega_{y} ^{\bar{y} \bar{r}} =  - \left( e^{a|r| + u/2} \right)'; \hspace{5 pt}  }\nonumber\\
& & \Omega_{z} ^{\bar{z} \bar{r}} =  - \left( e^{a|r| - u} \right)'; \hspace{5 pt} \Omega_{x } ^{\bar{x} \bar{t }} = \Omega_{y} ^{\bar{y} \bar{t}} =   \left( e^{ u/2} \right)^{.}; \hspace{5 pt} \Omega_{z} ^{\bar{z} \bar{t}} = \left( e^{-u}\right)^{.}.
\end{eqnarray}
From the action  given by Eq. (\ref{fermion-action}) we derive the  equation of motion 
\begin{equation} \label{fermion-eq-motion}
\left( \Gamma ^{\mu} D_{\mu} + \Gamma ^{r} D_{r} + \right) \Psi(x^{M}) = 0,
\end{equation}
\noindent and we assume  the standard chiral decomposition for the wave function  
\begin{equation}\label{fermion-wave-separ}
\Psi(x^{\nu}, r) = \psi_L(x^{\nu})\lambda (r) + \psi_R(x^{\nu})\rho(r).
\end{equation}

Assuming, once again, that the frequency of fermions on the brane are much smaller than the standing waves frequency one can substitute the oscilatory function in Eq. (\ref{fermion-eq-motion}) by its related time average versions. Using Eqs. (\ref{average-u}), (\ref{average-exp}) and (\ref{average-deriv-exp}), and the relations on Eqs.  (\ref{gamma-matr}), (\ref{spin-conx-nonvan}) and (\ref{fermion-wave-separ}), and by its turn, applying in Eq. (\ref{fermion-eq-motion}), we can found equations for  the \textit{r-dependent} factors of the left and right fermions. The equations for the extra dimension variable are, respectively \cite{Merab2012},
\begin{equation}\label{fermion-solution-lambda}
\lambda'' - \left[5a\; sgn(r) - \frac{P'}{P} \right]\lambda' + \left[ 4a\delta(r) + 6a^2 - 2a\; sgn(r) \frac{P'}{P} - P^2e^{-2a|r|} \right]\lambda = 0,
\end{equation}
\noindent and
\begin{equation}\label{fermion-solution-ro}
\rho(r)D = e^{a|r|}\frac{\sigma^iP_i(r)}{P^2(r)}\left[ 2a\; sgn(r) + \partial_r \right]\lambda(r)L.
\end{equation}
where $\sigma^i$ ($i = x, y, z$) represent the standard  $2\times2$ Pauli matrices.
\noindent Here $P(r) = P_x^2 + P_y^2 + P_z^2$ is interpreted as a `` \textit{r-dependent moment}'' for the spinor field  and the components are given, respectively, as 
\begin{equation*}
P_x(r) = \left[ I_0(|C_1|Q(r)/2) - 1 \right]p_x,
\end{equation*}
\begin{equation}
P_y(r) = \left[ I_0(|C_1|Q(r)/2) - 1 \right]p_y,
\end{equation}
\begin{equation*}
P_z(r) = \left[ I_0(|C_1|Q(r)) - 1 \right]p_z,
\end{equation*}
\noindent where the function $Q(r)$ is the standing wave solution given in Eq. (\ref{disp-vec-partic-sol}), $L$ and $D$ are  constant 2-spinors.
It is not possible to solve Eq. (\ref{fermion-solution-lambda}) algebraically. However, as for the bosonic fields considered above, we only need to consider its asymptotic form close and far from the brane.   In the first case, close to the brane the results for left and right fermions are, respectively
\begin{equation}\label{left-fermion-sol-close-aprox}
\lambda(r)|_{r\rightarrow \pm 0} = Be^{-2a|r|},
\end{equation} 
\noindent where $B$ is a constant, and
\begin{equation}\label{right-fermion-sol-close-aprox}
\rho(r)|_{r\rightarrow \pm 0} = 0.
\end{equation}
In the second case, far from the brane, the respective asymptotic values for    left and right fermions are
\begin{equation}\label{left-fermion-sol-far-aprox}
\lambda(r)|_{r\rightarrow \pm \infty} \sim e^{-3a|r|},
\end{equation} 
\noindent and
\begin{equation}\label{right-fermion-sol-far-aprox}
\rho(r)|_{r\rightarrow \pm \infty} \sim e^{-2a|r|}.
\end{equation}

Relations in Eq. (\ref{left-fermion-sol-close-aprox}) and Eq. (\ref{left-fermion-sol-far-aprox}) show that left fermion function has a maximum on the brane, $\lambda(r)|_{r = 0} = B$ and is convergent far from the brane which assures localization of leff fermion zero mode on the brane.  By other side, Eq. (\ref{right-fermion-sol-close-aprox}) shows that right fermion zero mode is not localized on the brane.

\section{Remarks and conclusions} \label{section5}
In this work, we have considered a five-dimensional braneworld built within the Lyra manifold. Particularly, we have adopted an anisotropic brane obtained from standing wave solutions.

After establishing the setup, we proceed with the analysis of how likely is the Lyra standing wave scenario, from the perspective of matter field localization. Our motivation in choosing such model arises from the previous results in field localization on geometrical braneworlds. Particularly, the vector field the zero mode cannot be localized on the Weyl brane without an additional localization procedure based on the interaction of vector field and Weyl scalar.

Firstly, we have considered the localization of the scalar field.  From the five-dimensional action with the kinetic term coupled to gravity we have obtained the equations of motion corresponding. Is important to mention that the scalar field on the Eq. (\ref{scalar-action}) is not the stuff the brane is made off, like the dynamically generated thick brane models proposed early.  From the evaluation at $r \rightarrow 0 $ and $r \rightarrow \infty$, the wave function resulting from the equations of motion, we can assure the localization to the scalar field zero mode. Aditionally we have analysed the role of a time dependent mass in the mechanism of filed localization, in our specific case. The time dependent mass was introduced in the scalar field equation through a nonminimal coupling. The results shows that in this case the time dependent mass is important in the process of field localization. 

We pay particular attention to the behaviour of the gauge field on the standing wave context because usually, to obtain the zero mode localization, some authors have included new couplings on the lagrangian. In the present manuscript, we have not considered any additional localization mechanism. The zero mode localization is achieved directly by the standing wave geometry. Thus, by the equations of motion of the gauge field component on the extra dimension, we find that the wave function for zero mode has a maximum on the brane and falls of far from the brane. On this way, we can assure the zero mode localization.  

Turning our attention to the fermionic field, we firstly connect the Dirac matrices on flat and curved spacetime using the standing wave metric. Extracting the part dependent on the extra dimension of the equations of motion we are unable to find analytical solutions. However, we take into account the asymptotic form close and far from the brane of such solutions.  The respective asymptotic values for left and right fermions show that left fermion function has a maximum on the brane and is convergent far from the brane.  The right solution assumes zero value at r=0.  From this behavior we can conclude that there is a zero mode localized for the left fermion, but the right fermion zero mode is not localized on the brane.

\section*{Acknowledgments}
\hspace{0.5cm}The authors would like to thank the Funda\c{c}\~{a}o Cearense de Apoio ao Desenvolvimento Cient\'{\i}fico e Tecnol\'{o}gico (FUNCAP)(PNE-0112-00061.01.00/16), and the Conselho Nacional de Desenvolvimento Cient\'{\i}fico e Tecnol\'{o}gico (CNPq) for grants Nos. 307688/2016-0 (WTC), 312356/2017-0 (JEGS) and 308638/2015-8 (CASA). 



\begin{thebibliography}{99}




\bibitem{Randal1999}L. Randall and R. Sundrum, Phys. Rev. Lett. 83 (1999) 3370.

\bibitem{Arkani2001}N. Arkani-Hamed, S. Dimopoulos and J. March-Russel, Phys. Rev. D 63 (2001) 064020.

\bibitem{Arkani2000}N. Arkani-Hamed, S. Dimopoulos, G. Dvali and N. Kaloper, J. High Energy Phys. 012 (2000) 010.

\bibitem{Chen2000} J. W. Chen, M. A. Luty and E. Ponton, J. High Energy Phys. 09 (2000) 012.

\bibitem{Khoury2001}J. Khoury, B. A. Ovrut, P. J. Steinhardt, and N. Turok, Phys. Rev. D 64 (2001) 123522.

\bibitem{Tao2017} Tao-Tao Sui, Li Zhao, Yu-Peng Zhang, Qun-Ying Xie, Eur. Phys. J. C77, 6 (2017) 411.

\bibitem{Oda2000a} Oda, Phys. Lett. B 496 (2000) 113.

\bibitem{Oda2000b} I. Oda, Phys. Rev. D 62 (2000) 126009.

\bibitem{Gherghetta2000} T. Gherghetta and M. E. Shaposhnikov, Phys. Rev. Lett. 85 (2000) 240.

\bibitem{Giovannini2001} M. Giovannini, H. Meyer and M. E. Shaposhnikov, Nucl. Phys. B 619 (2001) 615.

\bibitem{Kehagias2001} A. Kehagias and K. Tamvakis, Phys. Lett. B 504 (2001) 38.

\bibitem{nosso4} W.T. Cruz, A. R. P. Lima  and C. A. S. Almeida, Phys. Rev. D \textbf{87}, (2013) 045018.

\bibitem{nosso5} W.T. Cruz, R.V. Maluf and C.A.S. Almeida, Eur. Phys. J. C \textbf{73}, (2013) 2523.

\bibitem{Torrealba2010} R. S. Torrealba, Phys. Rev. D 82 (2010) 024034.

\bibitem{Gogberashvili2011}  M. Gogberashvili, P. Midodashvili and L. Midodashvili, Phys. Lett. B 702 (2011) 276.

\bibitem{Silva2011} J. E. G. Silva and C. A. S. Almeida, Phys. Rev. D 84 (2011) 085027.

\bibitem{Gogberashvili2012} M. Gogberashvili, P. Midodashvili and L. Midodashvili, Phys. Lett. B 707 (2012) 169.

\bibitem{Sousa2012} L. J.S. Sousa, W. T. Cruz, C.A.S. Almeida, Phys. Lett. B 711 (2012) 97.

\bibitem{Merab2012}M. Gogberashvili, JHEP 1209 (2012) 056.

\bibitem{Sousa2014} L. J. S. Sousa, W. T. Cruz, C .A. S. Almeida, Phys. Rev. D 89 (2014) 064006

\bibitem{Sousa2014} L. J. S. Sousa; C. A. S. Silva, D. M. Dantas, C. A. S. Almeida, Phys. Lett. B 731 (2014) 64

\bibitem{Cruz2016} W. T. Cruz, L. J. Sousa, R. V. Maluf, C.A.S. Almeida, Phys. Lett. B 730 (2014) 314.

\bibitem{Cruz2016} W.T. Cruz; R. V. Maluf ; L. J. S.  Sousa; C. A. S. Almeida,  An. of Phys., 364 (2016), 25.

\bibitem{Merab2009} M. Gogberashvili, D. Singleton, Mod. Phys. Lett. A25 (2010) 2131.

\bibitem{Sen-1972} D. K. Sen and J. R. Vanstone, J. of Math. Phys. 13 (1972) 990. 

\bibitem{Sen-1971} D. K. Sen and K. A. Dunn, A Scalar, J. of Math. Phys. 12 (1971) 578. 


\bibitem{HALFORD-1970} W. D. Halford, Australian Journal of Physics, vol. 23, (1970) 863. 

\bibitem{Casana2005} R. Casana, C. A. M. de Melo, B. M. Pimentel,  Braz. J. of Phys., 35(4B) (2005) 1151.


\bibitem{Gad2015} R. M. Gad, R. M., Inter. J. of Th. Phys., 54(8) (2015) 2932.

\bibitem{Darabi2015} F. Darabi, Y. Heydarzade, F. Hajkarim,  Can. Jour. of Phys., 93(12) (2015) 1566.

\bibitem{Shchigolev2015} V. K. Shchigolev, E. A. Semenova, (2015). Inter. Jour. of Adv. Astr. 3 (2) (2015) 117.

\bibitem{Saadat2016} H. Saadat, Inter. Jour. of Theor. Phys. 55(5) (2016) 2364.

\bibitem{Shchigolev2017} V. K. Shchigolev,  D. N. Bezbatko, Gravit. Cosmol. 24 (2018) 161.

\bibitem{Rahaman2001} F. Rahaman, J. K Bera, Int. Jour. of Mod. Phys. D 10(05) (2001) 729.




\bibitem{Mohanty2009} G. Mohanty, R. R. Sahoo, B. K. Bishi, Astrophys. and Sp. Sci., 319(1) (2009) 75.



\bibitem{Merab2011} M. Gogberashvili, A. Herrera-Aguilar and D. Malagón-Morejón, Class. Quantum Grav. 29 (2012) 025007.

















\end{thebibliography}
\end{document}